# Real-World Problem-Solving Class is Correlated with Higher Student Persistence in Engineering

Nathan Davis[1], Eric Burkholder[1,2]

[1]Department of Physics, Auburn University, Auburn AL 36849
[2]Department of Chemical Engineering, Auburn University, Auburn AL 36849

**Abstract:** Student persistence in science, technology, engineering, and mathematics (STEM) has long been a focus of educational research, with both quantitative and qualitative methods being used to investigate patterns and mechanisms of attrition. Some studies have used machine learning to predict a student's likelihood to persist given measurable classroom factors and institutional data, while others have framed persistence as a function of a student's social integration in the classroom. While these methods have provided insight into broader underlying patterns of attrition in STEM, they have not investigated class structures or teaching methods that promote persistence. In this study we explore how a research-based instructional format for an introductory calculus-based physics class using real world problem-solving (RPS) was correlated with higher persistence for students at a large research-intensive university. We found that the one-year persistence rates for the RPS course were 74% (fall semester) and 90% (spring semester), while the lecture-based class had a persistence rate of 64% and 78%, respectively. In spring, the RPS persistence rate was significantly higher (p=0.037). The RPS also had higher final grades and larger learning gains than the lecture-based class despite lower scores on a physics diagnostic test. We also note that the higher rates of persistence were not completely explained by higher final grades. This study motivates future work to understand the structural mechanisms that promote student persistence in introductory physics courses.

## Introduction

Persistence in science, technology, engineering, and mathematics (STEM) has been a long-standing issue in education research. Diminishing enrollment and high attrition rates have motivated leadership at many academic institutions to enact high-level institutional reform as well as examine social and motivational factors to understand and mitigate this growing concern [1,2]. However, little has been done to see how course structure in physics might influence persistence. In this study we provide an example of how using research-based teaching methods in an introductory calculus-based physics class can impact students' persistence in engineering one year after completing the course. While we do not yet have statistical power to build potential mediation or moderation models [3], our results show the potential for how modifying course structures can not only improve student learning and performance (as previously been demonstrated [4]) but also promote student persistence. As prior literature illustrates, the problem of student persistence is a complex one which can be studied through many lenses, and which depends on a multitude of (interacting) factors. Our aim with the present study is to provide a specific example of how course structures in physics may impact student persistence and motivate future work that focuses on *who* these structures impact most and *how*.

## Background

Student persistence has historically been viewed through models of institutional departure. Tinto's model of institutional departure framed the student progression through college as a "rite of passage" broken into three milestones: separation, transition, and incorporation. The separation stage is characterized as a student's diminished engagement in relationships with friends, family, and others prior to enrolling in college. Transition marks the point in which the student develops new relationships with members of the group they seek to be a part of (e.g., other university students). Lastly, incorporation is defined by the student's participation in social interactions commonly practiced by the group (university social events, etc.) as well as participating as an active member in the group. Tinto argued that a student's likelihood of persistence is largely governed by their integration and acceptance, both socially and academically, at each stage [5]. He posited that students facing academic difficulty, students struggling to resolve their educational and occupational goals, and those struggling to become or remain integrated within the intellectual and social environment of their institution were more likely to depart [5]. While Tinto's framework provided an initial means to examine the pathways to get a STEM degree, it lacked the resolution to examine how instructional methodologies, student demographics, and institutional types affect student attrition.

Though active learning has become more common in university classrooms because of its positive impacts on learning [6,7] there is relatively little research on how it affects student departure from STEM. Tinto saw the importance that the classroom played socially for a student's integration but did not examine how emerging instructional methodologies could directly affect student attrition [8]. To address these shortcomings, Braxton et al. conducted a longitudinal study and found that faculty classroom behavior and active learning strategies influenced students' persistence or departure decision by directly affecting core constructs of Tinto's model like social integration [9]. Curiously, this study suggested that group work in the classroom did not have a statistically significant effect on social integration. One explanation for this could have been that the study did not factor in current understanding on the complexities of student dynamics in the construction of effective groups to promote collaborative learning [10].

It is important to note the distinction between institutional departure and STEM departure. While foundational studies previously focused mostly on institutional departure, more recent work has focused on students departing *STEM* while remaining at the same institution. In *Talking About Leaving,* the authors found that students experience either a "push" factor – meaning issues resulting from prior or current college major selection that are motivating them to leave--or a "pull" factor which relates to the opportunities that attract students from their current major because of potentials in academic and professional possibilities [11]. However, there are also other considerations that may alter a student's choice such as poor teaching practices, which was cited by 90% of switchers as a primary reason for leaving along with curriculum design and grades [11]. Due to sample size limitations, we do not distinguish institutional and STEM departure in our analyses here. Future investigations with larger sample sizes may be able to distinguish whether active learning formats are associated with lower levels of institutional departure *and* STEM departure.

While Ref. [9] provided key insight into how active-learning strategies can influence student departure, there are important differences between the scope of that work and our focus here. First, Braxton et al. examined a student population that consisted exclusively of first-time, full-time freshmen students with no specified major field of study at a highly selective private research institution. Moreover, the focus of Braxton et al. work was how general active-learning strategies such as class discussions or group work could affect student institutional departure. Here, we focus on a specific active-learning format within a single course impacts engineering students specifically, highlighting the profound impacts that certain courses may have on *STEM* departure.

With continual advancements in data collection efforts by academic institutions, researchers now have an opportunity to develop quantitative models to examine what features are the best predictors of student persistence. With the emergence of machine learning as a more commonly used technique in educational research, modeling student pathways has revealed key insights into important governing features that predict student persistence in institutional data sets. For example, Stewart et al. used logistic regressions to show how standardized test scores and high school grades played a major factor in determining a student's pathway to getting a physics degree [12]. More recently, these techniques have started to be applied to looking at *who* is departing. For example, Bean used multiple regression and path analysis to determine that male and female students leave academic programs for different reasons. Whereas female students were more likely to stay in an academic program given the future benefits of gainful employment, male students were interested in personal development and freedom [13]. Others have developed predictive models to try to preemptively curb student departure early in their academic careers. One study applied discriminate analysis to predict the persistence of freshmen students and found that student-faculty relationships were strongly correlated with the students' decisions [14].

## Class Structure

The current study contrasts student outcomes following a standard introductory calculus-based physics class, which relied almost entirely on didactic lecture, with a real-world problem-solving (RPS) class which is highly interactive, focuses on teaching students real-world problem-solving skills, and is designed based on the theory of deliberate practice. The precise nature of the assignments and class format of the RPS course are given in greater detail in Ref. [15]. We drew data from two iterations of RPS (taught by E.B.), and three iterations of lecture-based sections of the same course across two semesters. Lecture (1) was taught by the same instructor in both semesters, while Lecture (2) represents a second instructor who taught another lecture-based course covering the same material. We summarize the differences in assessment and grading structures between the different instructors in Table I below.

*Table 1*: *Comparison of grading structure for the Fall 2021 and Spring 2022 Lecture and Real-World Problem-Solving (RPS) courses. RPS(1) represents the Real-World Problem-Solving Course taught by a single instructor both semesters where the number in parentheses next to the RPS and Lecture labels corresponding to the instructor for the course. The number in parentheses next to the percentage for In-Class Exams highlights the number of assessments given during the duration of the course for that given category.*

| **Fall 2021** | **Homework** | **Recitation** | **Lab** | **In-class Exams** | **Final Exam** | **Attendance/In-class worksheets** |
|---|---|---|---|---|---|---|
| Lecture(1) | 10% | 10% | 10% | 45% (3) | 25% | - |
| RPS(1) | 25% | - | 10% | 40% (4) | 15% | 10% |
| **Spring 2022** | | | | | | |
| Lecture(1) | 10% | 10% | 10% | 45% (4) | 25% | - |
| Lecture(2) | 10% | 5% | 15% | 35% (3) | 30% | 5% |
| RPS(1) | 30% | - | 12% | 36% (4) | 12% | 10% |

The lecture-based sections used an automated online homework system which provides binary grades (e.g., right, or wrong) on standard textbook-style problems. The homework assignments for the RPS course were developed by E. B. and employed a mix of conceptual explanation problems and real-world problems that were scaffolded with the assistance of a problem-solving template [16]. See Fig. 1 for an example. The RPS course dropped the lowest homework grade in both semesters. Another distinction between the lecture and the RPS course is the format and stakes of in-class assessments. For the RPS course (in the Fall) the instructor implemented a "two-stage" quiz [17] where for the first 30 minutes of class students solved real-world problems individually. Students turned in the individual solutions and then formed ad-hoc groups and submitted a group solution to the same problems. Ninety percent of the total grade was based on the individual portion and 10% was based on the group portion (but the final score could never be lower than a student's individual score).

*Figure 1: Template from the RPS course, adapted from Ref. [16].*

In the Spring, the RPS class used similar (but not identical) exam questions, but instead of a two-stage format, students were allowed to submit quiz "reflections" for 50% of their missed points back. These reflections asked students to provide a detailed explanation of what mistakes they made and how they might avoid those mistakes in the future; students were allowed to consult with peers or teaching assistants (TAs) for this part. Finally, quizzes in the RPS course in both semesters were entirely open resource (including books, class notes, homework solutions, online resources etc.). In both semesters, the lowest quiz grade was dropped. The lecture courses implemented standard closed-note, closed-book exams that were taken individually during class time. The homework and in-class assessment structure were the same in the Fall and Spring for the lecture-based courses.

The recitation sections in the lecture-based courses consisted primarily of weekly quizzes with little-to-no review of material. The recitations in the RPS course were devoted to review of material and practice problems, but how the review was structured was left to individual teaching assistants. The lab activities were identical between all sections and consisted primarily of "cookbook" activities hypothesized to reinforce content. (There is ample evidence in the literature that these are not beneficial to student learning [18], but the design of these activities was beyond the control of the RPS instructor.) All class sections took a common final exam which was written by other faculty in the department not currently teaching the course. These exams consisted of 10 conceptual multiple-choice questions, 10 short-calculation textbook-style questions (also multiple-choice) and 5 free-response, textbook-style questions. For an example of a textbook problem compared to a real-world problem, see Table .

*Table 2: Textbook problem (left) and authentic problem (right).*

| Textbook Problem (Serway, et. al, 2004 6$^{th}$ Ed., pg. 379, problem 12) | Real-World Problem |
|---|---|
| A 20.0-kg floodlight in a park is supported at the end of a horizontal beam of negligible mass that is hinged to a pole. A cable at an angle of 30.0° with the beam helps to support the light. Find (a) the tension in the cable and (b) the horizontal and vertical forces exerted on the beam by the pole. | An easier alternative to a traditional push-up is the incline push-up, where your hands are on an inclined surface above your feet. Calculate how much easier this is than a regular pushup. You should come up with an answer in terms of variables that could be easily measured (e.g., weights, distances) |

In-class, the lecture-based classes followed a standard lecture format that consisted of students listening to the instructor for the majority of the class while being presented the material of interest. Sometimes the lecture-based class would use clicker questions for purposes of recording attendance or course participation, but no other active learning strategies were implemented, and the courses did not typically follow best practices for use of clickers [19]. The RPS class time was mostly spent on students' usage of scaffolded worksheets while solving real-world physics problems [15]. Course time was mostly spent on scaffolded group activities with feedback from learning assistants (LAs). However, the instructor would periodically bring all the groups together to review concepts and ask clicker questions (using a peer instruction format) to assess student understanding and transfer. Note that the Fall 2021 RPS course was in a theater-style lecture hall, whereas the Spring 2022 iteration was in a SCALE-UP classroom [20].

## Methodology

To measure the differences in persistence for students in engineering we collected information on students' majors while they were taking the course, and then again one year later, from institutional records. We used a one-year window as it has been shown that approximately 50% of attrition from STEM typically happens within the first year. [12].Data on students' prior physics and math knowledge were collected using a survey that was administered at the beginning and end of each course (in the first and last recitation sections of the semester). This was a validated physics diagnostic [21] that tested relevant math and physics concepts such as vector manipulations, basic calculus questions, questions about forces and motion, etc.  All students enrolled in the courses were asked to complete the survey as a part of the lab grade, but the answers were not graded on correctness.  After filtering students who did not complete the survey, completed the survey in less than 10 minutes (average time was 35 minutes), or did not consent to have their data used for research, we had a sample size of N=782 students. The breakdown of the sample size per semester and class type can be found in Table 3. Note that there were many more students in the lecture-courses in the Spring semester as that is "on-sequence" for engineering students. Students in the Fall semester are typically students with Advanced Placement (AP) Calculus credit or students who had to delay taking the course (for example, because they were not calculus-ready their first semester). The RPS sample size in the Spring is small because it was taught in a SCALE-UP classroom rather than a lecture hall [20]. Finally, 94% of students were engineering majors at the beginning of the class.

*Table 2: Sample size breakdown of the number of students for each respective class type and semester.*

|  | Fall | Spring |
|---|---|---|
| **Lecture** | 110 | 491 |
| **Real-World Problem-Solving** | 129 | 52 |

Due to students dropping the course after the first week, or not taking the post-semester assessment because the lowest lab grade was dropped, ~55% of the students had full information recorded. However, because the data contained a large number of variables for each of the 782 students, the total fraction of missing data was only 10.4%. Best practices for handling missing data suggest that this is beyond the acceptable threshold of 5%, under which listwise deletion is acceptable [22]. To rectify this, we used Predictive Mean Matching (PMMN) using 20 nearest neighbors and a gaussian as our perturbation from the Python(3.88) statsmodels (0.12.2) imputation mice package along with the pandas(1.2.4) package , which estimates values for missing features in the data set by matching with the observed values from other students who did complete it. Following the imputation of the data we then converted all continuous variables to z-scores (such that they were in units of standard deviations).

After preprocessing the data, we conducted t-tests to determine if the persistence rates and diagnostic scores across course-type and semester were statistically different. We then used logistic and linear regression analysis to test the strength of the relationships between potential explanatory variables like diagnostic scores and semester on the outcomes of interest. The models tested for this analysis were:

$$logit(p) \sim \beta_0 + \beta_1(\text{ProblemSolving}) + \beta_2(\text{Semester}) \quad (1)$$

$$logit(p) \sim \beta_0 + \beta_1(\text{ProblemSolving}) + \beta_2(\text{Semester}) + \beta_3(\text{Final Score}) \quad (2)$$

$$\text{Final Score} \sim \beta_0 + \beta_1(\text{PreDiagnostic}) + \beta_2(\text{ProblemSolving}) + \beta_3(\text{Semester}) \quad (3)$$

$$\text{PostDiagnostic} \sim \beta_0 + \beta_1(\text{PreDiagnostic}) + \beta_2(\text{ProblemSolving}) + \beta_3(\text{Semester}) \quad (4)$$

For Models (1) and (2) $logit(p) = \ln(\frac{p}{1-p})$ where p is the probability to persist. Model (1) tests whether persistence in the RPS course was higher (as indicated by the size of $\beta_1$) while controlling for differences between the overall persistence rates in the two semesters ($\beta_2$). Model (2) then tests whether the persistence rate in the RPS course was still larger when controlling for the correlation between course grade and persistence ($\beta_3$). Models (3) and (4) investigate how course type and semester can affect performance and learning while controlling for pre-diagnostic scores. For model (3) $\beta_0$ represents the average final course score for students with an average pre-diagnostic score in the lecture course in the fall semester. $\beta_1$ is the effect-size or correlation that the pre-diagnostic has on final course score, and $\beta_2$ and $\beta_3$ represent the difference in course grade by course and semester (respectively) while holding pre-diagnostic scores constant. The interpretation of model (4) is identical, but with the coefficients reflecting differences in post-diagnostic as the outcome.

## Results

We found that the percentage of students who remained engineering majors one year after completing the RPS course was significantly higher than the lecture-based courses in the Spring semester (p = 0.037; Table 3), though the persistence rates in the spring semester were higher in both course formats compared to the fall (p = 0.0015 for Control, p = 0.0107 for RPS). The persistence rate in the RPS course was also higher than the control in the fall, but this difference was only marginally significant (p = 0.071). However, we found that students in the RPS course on average had lower pre-diagnostic scores (t = -5.07, p < 0.001).

*Table 3: Persistence rates and their statistical significances foreach corresponding class type and semester.*

|  | Fall | Spring | p-value |
|---|---|---|---|
| **Lecture** | 64% | 78% | 0.0015 |
| **Problem-Solving** | 74% | 90% | 0.017 |
| **p-value** | 0.071 | 0.037 | - |

We next visualized the pathways of students from each course and semester using an Alluvial Plot (more commonly known as a Sankey diagram in physics; Figure 2). Most students start as engineering majors in both semesters, but there are some other STEM majors in the courses as well (typically physics, chemistry, or mathematics). Visually, it appears that the difference in attrition in the fall semester is from more students switching into non-STEM majors in the control course, rather than leaving the university or moving to another STEM major. The trends are similar in spring, though due to the small number of students leaving, we cannot make strong claims about the different non-engineering destinations of students across the two courses and semesters.

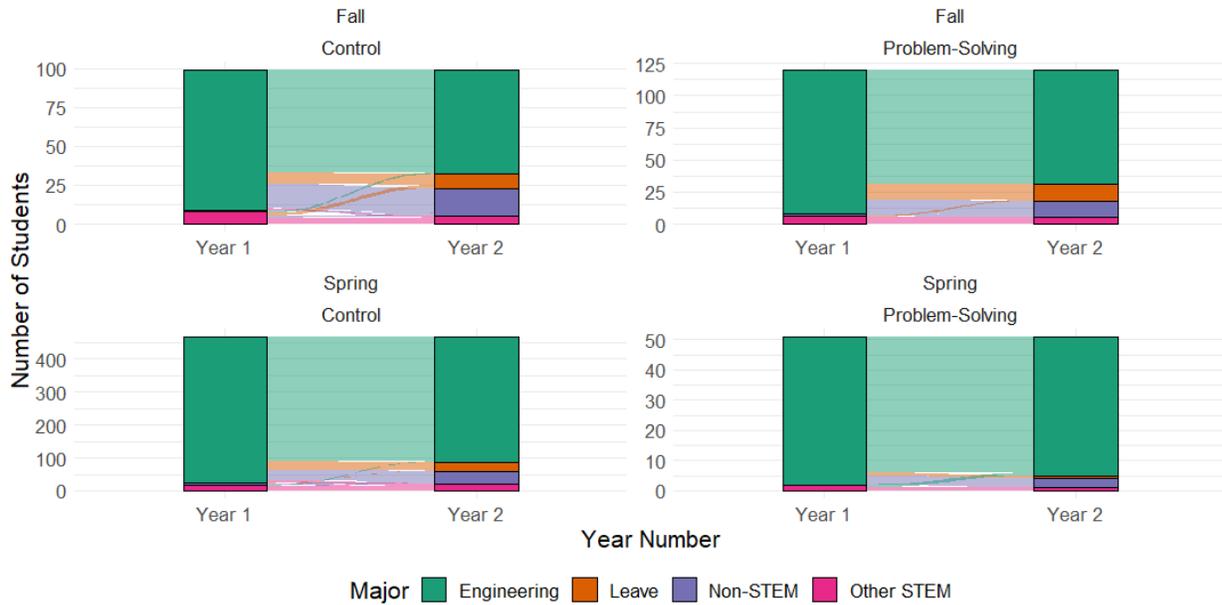

*Figure 2: Alluvial Diagram illustrating the different pathways students, from an introductory calculus-based physics course using either a traditional lecture format or real-world problem-solving format, take one year after completing the course.*

We next used a stepwise multiple logistic regression to determine the effect-sizes while controlling for the differences between semesters and final course grades (results in Table 5). We found that students in the RPS class were 1.9 times more likely to stay in engineering than in the lecture-based class (95% Confidence Limits [C.L.] = 1.2 – 3.04). We also found that, across both classes, students in the Spring were 2.2 (95% C.L. = 1.48 – 3.28) times more likely to stay in engineering. When controlling for final course grade, we found that the RPS students were 1.52 (95% C.L. = 0.94 – 2.53) times more likely to persist than their peers in the lecture-based courses who earned the same final grades. While the correlation is not significant at the p = 0.05 level when controlling for final grade, the effect-size is moderate. In the spring semester, the odds of persistence were factor of 1.63 (95% C.L. = 1.06 – 2.48) times higher compared to the fall semester (p=0.0253). Lastly, we also found that for a one standard deviation increase in the final grade, the odds of persistence increase by a factor of 1.80 (95% C.L. = 1.51-2.18) illustrating that students with higher final scores had higher odds of persisting 1 year after finishing the course.

*Table 5: Stepwise Logistic Regression Model (Persistence) where the confidence limits (C. L.) correspond to the calculated odds ratio or the exponentiated beta values. The $\beta_k$'s in this table correspond to the $\beta_k$'s outlined in the Methodology section for models (1) and (2).*

|  | Model | | | |
|---|---|---|---|---|
| **Step 1** | | | | |
| **Predictors** | $\beta_k$ | $e^{\beta_k}$ | 95% C.L. | p-value |
| Intercept | 0.50 | 1.65 | 1.15-2.35 | 0.00632 |
| Problem-Solving | 0.64 | 1.90 | 1.20-3.04 | 0.00706 |
| Semester | 0.79 | 2.20 | 1.48-3.28 | 9.75x10<sup>-5</sup> |

**Step 2**

| Predictors | $\beta_k$ | $e^{\beta_k}$ | 95% C.L. | p-value |
|---|---|---|---|---|
| Intercept | 0.82 | 2.27 | 1.55-3.39 | 3.78x10[-5] |
| Problem-Solving | 0.42 | 1.52 | 0.94 -2.53 | 0.0918 |
| Semester | 0.49 | 1.63 | 1.06-2.48 | 0.0253 |
| Final Score | 0.59 | 1.80 | 1.51-2.18 | 2.49x10[-10] |

Though all students in the Spring were more likely to persist, we found that students who enrolled in the problem-solving course during either the fall or spring semester had higher odds of persistence. We also found that a sizeable portion of this difference could be explained by RPS students receiving higher grades in the course. Indeed, Table 6 indicates that students in the RPS course received course grades 0.52 standard deviations larger than in the traditional lecture course while controlling for pre-diagnostic (which is typically strongly correlated with course grades in the lecture-based course; [23]).

To determine whether these increased grades were simply inflated due to different assignment structures, we examined student learning on the post-diagnostic test. We found that students in the RPS course scored 0.22 standard deviations higher on the post-diagnostic (95% C.L. = 0.07 – 0.37) than students in the lecture course with the same levels of incoming physics knowledge. In summary, the RPS students were learning more, earning higher grades, and persisting at higher rates than their peers in the lecture courses.

*Table 6*: Linear Regression Models where the respective dependent variables are Course Final Score and Post Diagnostic with 95% confidence limits (C.L.) corresponding to the calculated $\beta_k$. The $\beta_k$'s in this table correspond to the $\beta_k$'s outlined in the Methodology section for models (3) and (4).

| | Models | | |
|---|---|---|---|
| **Dependent (Final Score)** | | | |
| **Independent** | $\beta_k$ | 95% C.L. | p-value |
| Intercept | -0.42 | (-0.69)-(-0.27) | 4.42x10[-8] |
| Pre-Diagnostic | 0.37 | 0.31-0.44 | <2.0x10[-16] |
| Problem-Solving | 0.52 | 0.35-0.69 | 6.22x10[-9] |
| Semester | 0.43 | 0.27-0.59 | 1.63x10[-7] |
| **Dependent(Post Diagnostic)** | | | |
| **Independent** | $\beta_k$ | 95% C.L. | p-value |
| Intercept | -0.42 | (-0.55)-(-0.29) | 4.72x10[-10] |
| Pre-Diagnostic | 0.54 | 0.48-0.60 | <2.0x10[-16] |
| Problem-Solving | 0.22 | 0.07-0.37 | 0.0054 |
| Semester | 0.53 | 0.39-0.67 | 2.48x10[-13] |

## Discussion

We observed a higher level of one-year persistence of engineering of students in the RPS course. In particular, the RPS course had students with lower pre-test diagnostic scores who still earned higher grades and saw larger learning gains than in the lecture-based courses. This is consistent with prior data [15] suggesting that diagnostic score was not as strongly correlated with success in the RPS course [15]. The difference in persistence rates between Fall and Spring is likely due to the student populations. Students who take the introductory calculus-based physics course in spring are on track with their target program versus those who take the course in fall who are not. This too has been shown to be largely correlated with persistence especially from a financial and timing perspective for scheduling and availability, for future course work [24].

There are a number of limitations of the work we present here. First, we only examined the one-year persistence in this student population due to the limitations of the data available to us. We note that even though students may have maintained their majors one year after completing the course, it is possible that they can leave their programs at similar rates several years later. We are also limited by sample size in the current investigation. This again is for practical reasons, in that the RPS course is not yet offered on a regular schedule. To build up sufficient statistical power to make stronger claims about where these students go, we would need to conduct an in-depth longitudinal study with consistent delivery of the RPS course. This study is underway as the RPS course is now being offered annually in the fall semester. (We chose the fall semester as the overall retention rates are lower, and thus the potential impact on students would be greater.)

Similarly, we do not yet have sufficient quantitative data to test possible mechanisms of persistence across the two courses. In a future study we plan to use data from a social network survey we administered in Fall 2023 and perform Social Network Analysis (SNA) to investigate social mechanisms that may be contributing to what we are seeing in this study. Similarly, we have begun collecting data on motivational factors like test-anxiety, sense-of-belonging, and self-efficacy, which are also theorized to contribute to retention.

Perhaps most notably, we do not have sufficient data on *who* is departing following these introductory physics courses. Recent studies have shown that persons excluded because of their ethnicity or race (PEERs) [25], or first-generation students are often excluded from STEM degree attainment because of a lack of prior academic opportunities [26]. Other studies have examined the correlation of specific social and psychological factors like belonging, self-efficacy, and identity which are large contributors for a growing gender disparity across STEM degree attainment [27]. Motivated by these previous studies, researchers are attempting to quantify or develop predicative models of persistence to characterize how features like demographics, measures of prior academic achievement, or other social/psychological factors may have large contributing effects on student departure from STEM.

## Conclusion

In this study we examined how a research-based instructional format focused on real-world problem-solving affected the persistence rates of engineering students one year after completing the course. We found that those who participated in the RPS course had a larger likelihood of persisting when compared with the lecture-based course. Moreover, students from RPS course also had higher course grades, even for students who scored lower on a physics diagnostic test, and larger learning gains when compared to the lecture-based course. These outcomes have motivated future study to understanding the underlying mechanism of the RPS course to further promote student persistence in introductory physics.